\documentclass[onecolumn,manuscript,preprint,aps,pra]{revtex4}

\usepackage{bm,amsmath,amssymb,stmaryrd} \usepackage{graphicx}
\usepackage{amssymb}
\usepackage{epsfig}
\usepackage[english]{babel}
\usepackage{latexsym}
\usepackage{graphics}
\usepackage{subfigure}
\usepackage{epsfig}
\usepackage{graphicx}
\usepackage{dcolumn}
\usepackage{amsmath}

\def\PRA{{\textit{Phys.~Rev.~A}} }

\def\JPB{{\textit{J.~Phys.~B:At. Mol. Opt. Phys.}} }
\def\CPB{{Chin.~Phys.~B} }
\def\CPL{{Chin.~Phys.~Lett.} }
\def\PRL{{\textit{Phys.~Rev.~Lett.}} }

\def\NP{{\textit{Nat. Photon.}} }

\begin{document}


\title{Photoelectron angular distribution from the high-order above-threshold ionization process in IR+XUV two-color laser fields}

\author{Facheng Jin$^{1,2}$, Fei Li$^{1}$, Jing Chen$^{3\dag}$, Xiaojun Liu$^{4\sharp}$,Bingbing Wang$^{1,5*}$}

\address{$^1$Laboratory of Optical Physics, Beijing National Laboratory for Condensed Matter Physics, Institute of Physics, Chinese Academy of Sciences, Beijing 100190, China}
\address{$^2$Facualty of Science, Xi'an Aeronautical University, Xi'an 710077,  China}
\address{$^{3}$Institute of Applied Physics and Computational Mathematics, P. O. Box 8009, Beijing 100088, China}
\address{$^4$State Key Laboratory of Magnetic Resonance and Atomic and Molecular Physics, Wuhan Institute of Physics and Mathematics, Chinese Academy of Sciences, Wuhan 430071, China}
\address{$^5$University of Chinese Academy of Sciences, Beijing 100094,  China}
\date{\today}

\begin{abstract}
High-order above-threshold ionization (HATI) spectrum in IR+XUV two-color laser fields has been investigated. We found that the quantum features corresponding to the absorption of the XUV photon is well illustrated by a peculiar dip structure in the second plateau of the HATI spectrum. By the channel analysis, we show that the angular distribution of the spectrum is attributed to the coherent summation over contributions of different channels, and the dip structure in the spectrum is directly related to the absorption of one XUV photon of the ionized electron during the laser-assisted collision (LAC) with its parent ion in the two-color laser fields. Moreover, by employing the saddle-point approximation, we obtain the classical energy orbit equation, and find that the dip structure comes from the fact that the LAC is limited at a certain direction by the momentum conservation law as the electron absorbs one XUV photon during the collision, where the probability of the HATI gets its minimum value. Finally, we find that the interference pattern in the whole spectrum is attributed to the interference of different orbits at collision moments $t_0$ and $2\pi/\omega_1-t_0$ in the LAC process.
\end{abstract}


\maketitle


\par\noindent
$^{\dag}$chen$\_$jing@iapcm.ac.cn,
$^{\sharp}$xjliu@wipm.ac.cn, $^{*}$wbb@aphy.iphy.ac.cn

\section{INTRODUCTION}
The above-threshold ionization (ATI) has attracted much more interest and become one of the most active topics in strong-field atomic and molecular physics~\cite{Becker2002,Milo2006} since its first observation~\cite{Agostini1979}. In this process, an electron can absorb photons in excess of the minimum photon number necessary to overcome the ionization potential. It was found recently that, the very low energy spectrum of the ionized electron is attributed to the forward scattering of electron to its parent ion, i.e., the effect of the coulomb potential~\cite{Blaga2009,Quan2009,Xu2017} during the ionization. In general, the photoelectron spectrum may be classified by direct ATI part and high-order ATI (HATI) part, where the HATI spectrum is explained by the well-known three-step model~\cite{Corkum1993}: the electron is ionized through tunneling, and then accelerated, and finally  driven back to collide with parent ion and scatter elastically off in an intense infrared (IR) laser field, where the electron can obtain more kinetic energy in this process. The high-order ATI (HATI) process in an IR laser field has been extensively investigated and has been made great achievement. It is well known that plateau structure is a common phenomenon in the HATI process~\cite{Paulus1994,Paulus2003,Yu2017}. It was found that the position of the resonant structure of rare-gas atoms does not shift with the change of the laser intensity~\cite{Freeman1987,Hansch1997,Hertlein1997,Muller1999} and the resonantlike enhancement of molecules is attributed to the channel-closing mechanism in the HATI process~\cite{Quan2013,Wang2014,Wang2016}. The bicircular HATI processes of atom and molecule were discussed in detail~\cite{Milo12016,Milo12017}. Furthermore, it was proved that the angular distributions of atoms and molecules provide a sensitive perspective to further analyse the HATI process~\cite{Yang1993,Busulad2008,Kang2010,Paulus2000,Hasovi2015,Guo2009,Zille2017}, especially the laser-driven scattering process of an electron and the image of the molecular geometrical structure.

With the rapid development of laser technology~\cite{McNeil2010,Gallmann2012}, the combined IR and extreme ultraviolet (XUV) laser fields have offered an effective tool to investigate the electronic dynamics of atoms and molecules~\cite{Peng2015,Gallmann2012,Bengtsson2017}. For example, the control of the electron-correlation for double ionization~\cite{Hu2013,Liu2015a}, the molecular dissociative ionization~\cite{Kelkensberg2009,He2010} and the real-time observations of valence electron motion~\cite{Goulielmakis2004,Goulielmakis2010} were investigated in IR+XUV two-color laser fields. Furthermore, it was found that a sideband structure is presented in the photoelectron spectrum as well as the Auger electron spectrum~\cite{Kazansky2009,Kazansky2013,Kazansky2010}, which can be in favor of the investigation of the properties of pulses reversely. Also, these sidebands have been applied in investigating the molecular orbitals of O$_2$, H$_2$O and N$_2$~\cite{Leitner2015}. Additionally, the ATI spectrum presents a steplike structure when the XUV photon energy is much higher than the atomic ionization potential~\cite{Zhang2013,Wang2015} and the dependence of  ATI spectrum on polarization direction of two-color laser fields is identified~\cite{Liu2015}. Recently, the ATI process of atoms and ions in vortex XUV Bessel beam in the presence of a strong IR laser field was investigated~\cite{Seipt2016}. It was shown that the ATI photoionization spectra present seven different dichroism signals, which may open up avenues for future investigations of the interaction of atom and molecule with vortex laser field. Until now, the work of HATI process in two-color laser fields consisting on IR field and its second or third harmonic has been made much progress, such as the angular-dependent asymmetries of the HATI spectrum~\cite{lu2017}. More recently, the role of XUV laser field in the HATI was discussed~\cite{Zhang2017k}. It was demonstrated that electron absorbing XUV photons during the recollision process plays an important role in the high-energy region of the HATI spectrum. However, there are still few works on HATI in IR+XUV two-color laser fields, which needs more attention. There exist some open questions, for example, how the XUV laser influences the recollision process, and what the change of the HATI angular distribution will be.

 In this paper, we investigate the HATI process of an atom in IR+XUV two-color laser fields by employing the frequency-domain theory, where the energy of the XUV photon is much larger than the atomic ionization potential. It has been demonstrated that the frequency-domain theory can be facilitated to deal with the recollision process~\cite{Gao2000,Wang2007,Fu2001,Guo2009,Wang2010,Wang2012,Jin1,Jin2,Jin3}, such as HATI, high-order harmonic generation (HHG) and nonsequential double ionization (NSDI). Here we present the angular distribution of the HATI process with multiplateau structure, where there exists a dip structure in the second plateau. With the help of the channel analysis, one can see that the interference pattern comes from the contributions of different channels. Furthermore, the HATI can be treated as a two-step process: an ATI followed by a laser-assisted collision (LAC) process~\cite{Wang2007,Guo2009}. It is found that  the dip structure is attributed to the fact that one XUV photon is absorbed by the atom in the LAC process. Moreover, by using the saddle-point approximation, we obtain the classical energy orbit equation, which illustrates the formation of interference pattern. Atomic units are used throughout unless otherwise stated.

\section{THEORETICAL METHOD}

 The Hamiltonian of an atom in two-color linearly polarized laser fields is~\cite{Wang2007,Guo2009}
\begin{equation}\label{1}
  H=H_0+U+V,
\end{equation}
where $H_0=\frac{(-i\nabla)^2}{2}+\omega_1 N_1+\omega_2 N_2$ is the Hamiltonian of a free electron-laser system with the photon number operator $N_i$ for the laser field of frequency $\omega_i$ ($i=1,2$), $U$ is the interaction potential between an electron and the nucleus, and $V$ is the electron-photon interaction potential.

The transition matrix from the initial state $|\psi_i\rangle$ to the final state $|\psi_f\rangle$ for the HATI process is~\cite{Wang2007}
\begin{equation}\label{2}
  T_{\textmd{HATI}}=\langle \psi_f|U \frac{1}{E_f-H+i\epsilon} V | \psi_i  \rangle.
\end{equation}
Here, $| \psi_i\rangle=|\Phi_1\rangle\otimes |l_1\rangle\otimes |l_2\rangle$ with the initial energy $E_i=-I_p+(l_1+1/2)\omega_1+(l_2+1/2)\omega_2$, where $|\Phi_1\rangle$ is the ground state of an atom and $|l_j\rangle$ is the Fock state of the laser field with photon number $l_j$ for $j=1,2$. The final state $| \psi_f\rangle=| \psi_{\textbf{p}_f m_1 m_2}\rangle$ is the quantized-field Volkov state in the two-color laser field~\cite{Wang2012,Guo1992}
 \begin{equation}\label{3}
\begin{array}{l}
{|\psi_{\textbf{p}_f m_1 m_2}\rangle} =
V_e^{ - 1/2} \exp\{i[(\textbf{p}_f+u_{p_1}\textbf{k}_1+u_{p_2}\textbf{k}_2)\cdot \textbf{r}]\}\\
~~~~~~~~~ \times \sum\limits_{{j_1} =  - {m_1},{j_2} =  - {m_2}}^\infty \aleph_{j_1 j_2}(\zeta_f)^{*}\exp \{-i[j_1 (\textbf{k}_1 \cdot \textbf{r} +\phi_1)+j_2(\textbf{k}_2 \cdot \textbf{r} +\phi_2)]\}  \\
~~~~~~~~~  \times |m_1+j_1,m_2+j_2\rangle,
 \end{array}
\end{equation}	
with the energy of the final state $E_f=\textbf{p}^2_f/2+(m_1+1/2)\omega_1+(m_2+1/2)\omega_2+u_{p_1}\omega_1 +u_{p_2}\omega_2$. In the above, $V_e$ is the normalization volume, $\textbf{p}_f$ is the momentum of the ionized electron, $u_{p_j}=U_{p_j}/{\omega_j}$ with $U_{p_j}$ being the pondermotive energy, $\textbf{k}_j$ is the wave vector, and $\phi_j$ is the initial phase of the two laser fields with $j=1,2$. The term $\aleph_{j_1 j_2}(\zeta)$ in Eq.~(\ref{3}) is the generalized Bessel function, which can be written as~\cite{Guo1992}
\begin{equation}\label{4}
\begin{array}{l}
 {\aleph_{{j_1} {j_2}}}(\zeta) = \sum\limits_{{j_3}{j_4}{j_5}{j_6}} {{J_{ - {j_1} + 2{j_3} + {j_5} + {j_6}}}} ({\zeta_1}){J_{ - {j_2} + 2{j_4} + {j_5} - {j_6}}}({\zeta_2}) \\
 ~~~~~~~~~~~~~~ \times {J_{ - {j_3}}}({\zeta_3}){J_{ - {j_4}}}({\zeta_4}){J_{ - {j_5}}}({\zeta_5}){J_{ - {j_6}}}({\zeta_6}), \\
 \end{array}
\end{equation} 	
where
\begin{equation}\label{5}
\begin{gathered}
  {\zeta_{1}} = 2\sqrt {\frac{{{u_{p_1}}}}{{{\omega_1}}}}  {{\mathbf{\text{\textbf{p}}}}}\cdot {{\widehat{{\mathbf{\epsilon}}}}_1},~~~~~~~~~ {\zeta_{2}} = 2\sqrt{\frac{{{u_{p_2}}}}
{{{\omega _2}}}} {{\text{\textbf{p}}}} \cdot {{\widehat{{\mathbf{\epsilon}}}}_2}, \hfill \\
{\zeta_{3}} = \frac{1}{2}u_{p_1},~~~~~~~~~~~~~~~~~~~ {\zeta_{4}} = \frac{1}{2}u_{p_2}, \hfill \\
{\zeta_{5}} = 2\frac{{\sqrt {{u_{p_1}}{\omega _1}{u_{p_2}}{\omega _2}} }}
{{{\omega _1} + {\omega _2}}},~~~~~~ {\zeta_{6}} = 2\frac{{\sqrt {{u_{p_1}}{\omega _1}{u_{p_2}}{\omega _2}} }}{{{\omega _1} - {\omega _2}}}, \hfill \\
\end{gathered}
\end{equation}	
and $J_m(t)$ is the Bessel function of order $m$. Also, $\textbf{p}$ is the momentum of the ionized electron, ${{\widehat{{\mathbf{\epsilon}}}}_j}$ is the unit vector representing the polarization direction of laser field with $j=1,2$.

In the HATI process, an intermediate state can be expressed as $| \psi_{\textbf{p}_n n_1 n_2}\rangle$ with the energy $E_n={\textbf{p}^2_n}/2+(n_1+1/2)\omega_1+(n_2+1/2)\omega_2+u_{p_1}\omega_1 +u_{p_2}\omega_2$. By applying the completeness relation of the intermediate state, Eq.~(\ref{2}) can be rewritten as~\cite{Wang2007}
\begin{equation}\label{6}
\begin{array}{l}
 {T_{\textmd{HATI}}} = -i\pi \sum\limits_{\textbf{p}_n n_1 n_2}^\infty \langle \psi_{\textbf{p}_f m_1 m_2}|U|\psi_{\textbf{p}_n n_1 n_2}\rangle \langle \psi_{\textbf{p}_n n_1 n_2}|V|\psi_{i}\rangle \delta (E_f-E_i). \\
 \end{array}
\end{equation}
Here, we have noted that the term $\langle \psi_{\textbf{p}_n n_1 n_2}|V|\psi_{i}\rangle$ stands for the ATI process, where an electron is ionized directly by two-color laser fields, and the term $\langle \psi_{\textbf{p}_f m_1 m_2}|U|\psi_{\textbf{p}_n n_1 n_2}\rangle$ stands for the LAC process, where the ionized electron collides with the parent ion and scatters off by a laser-assisted collision. By using Eq.~(\ref{3}), the transition matrix for the HATI can be expressed as
\begin{equation}\label{7}
\begin{array}{l}
{T_{\textmd{HATI}}} =-\frac{i}{8\pi^{2}V_e^{1/2}} \sum\limits_{s_1 s_2} \sum\limits_{q_1 q_2} |\textbf{p}_n| [(u_{p_1}-s_1)\omega_1 + (u_{p_2}-s_2)\omega_2] e^{i(s_1 \phi_1+ s_2 \phi_2)} e^{i(q_1 \phi_1+ q_2 \phi_2)} \\
~~~~~~~~~~~~ \times \int \sin{\theta_n} d \theta_n d \varphi_n \aleph_{s_1 s_2} (\zeta_n) \aleph_{q_1 q_2} (\zeta_f-\zeta_n)  \phi_i(\textbf{p}_n) \int d\textbf{r} e^{-i\textbf{p}_f\cdot \textbf{r}} U e^{i\textbf{p}_n \cdot \textbf{r}}, \\
 \end{array}
\end{equation}	
where $s_1=l_1-n_1$ and $s_2=l_2-n_2$ are the number of photons absorbed from the first and second laser fields in the ATI process, $q_1=n_1-m_1$ and $q_2=n_2-m_2$ are the number of photons absorbed from the first and second laser fields in the LAC process. In above expression, $|\textbf{p}_n|=\sqrt{2[s_1 \omega_1 +s_2 \omega_2 -u_{p_1}\omega_1-u_{p_2}\omega_2-I_p]}$ is the magnitude of the electron momentum before collision, $\theta_n$ is the angle between the emission direction of the ionized electron and the laser polarization, $\varphi_n$ is the azimuth angle.

\section{NUMERICAL RESULTS}

In this section we will consider the HATI process of an atom in an IR+XUV two-color laser field. The atomic ionization threshold is $I_p=12.1$~eV. The frequencies of the IR and XUV laser fields are $\omega_1=1.165$~eV and $\omega_2=50\omega_1$, and their intensities are $I_1=I_2=1.0\times 10^{13}$~W/cm$^{2}$. The polarization directions of the two-color laser fields are the same and their initial phases are set to zero for simplify.

\subsection {The angular distribution of the HATI spectrum}

Figure~\ref{fig1}~(a) presents the HATI spectrum for different $\theta_f$ and (b) the angular distribution of HATI probability of an electron in the HATI process, where $\theta_f$ is the angle between the emission direction of the ionized electron and the laser polarization direction. One can see that the HATI spectrum, as well as, the angular distribution shows a multiplateau structure, where the probability of the first plateau is much higher than that of the second plateau by about six orders of magnitude. Furthermore, it is found that the probability distribution shown in Fig.~\ref{fig1}~(b) is symmetry about the line defined by $\theta_f=90^\circ$, where the width of the spectrum decreases (increases) with the value of $\theta_f$ as $\theta_f\leq90^\circ$ ($\theta_f>90^\circ$). Especially, the width of the spectrum shows narrowest around $\theta_f=90^\circ$. Also, we have noted that there is a narrow dip structure in the second plateau of the angular distribution. In what follows, we will explain the formation of the interference patterns and the emergence of the dip structure on the spectrum.

According to the previous investigations~\cite{Hu2013,Jin1,Jin2,Zhang2013,Liu2015}, we know that the XUV laser field plays a crucial role in the ionization process. Figure~\ref{fig2} shows the angular distribution of the atom absorbing one [Fig.~\ref{fig2}(a)] and two [Fig.~\ref{fig2}(b)] XUV photons in the HATI process. Comparing Fig.~\ref{fig1} with Figs.~\ref{fig2}(a) and \ref{fig2}(b), one can find that the first and second plateaus in Fig.~\ref{fig1} are separately attributed to the processes of atom absorbing one and two XUV photons. Therefore, the probability of the first plateau is much larger than that of the second plateau.

Based on the frequency-domain theory, the HATI can be decoupled into a two-step process, i.e., an ATI followed by an LAC. The atom may absorb XUV photons in both ATI and LAC processes. Hence we now define the channel as $(s_2,q_2)$, where $s_2$ and $q_2$ are the number of the XUV photons absorbed by an atom in the ATI and LAC processes, respectively. Firstly, we consider the first plateau of the spectrum where the electron absorbs one XUV photon in the HATI process. Figure~\ref{fig3} shows the angular distribution of the HATI spectrum for channels (0,1) [Fig.~\ref{fig3}(a)] and (1,0) [Fig.~\ref{fig3}(b)]. One can see that the channel (1,0) dominates almost all contributions to the first plateau. This can be understood as follows: since one XUV photon energy (i.e., 58.25~eV) is much larger than the atomic ionization threshold (i.e., 12.1~eV) in our calculation, the electron may have enough energy to be ionized by absorbing one XUV photon in the ATI process, as a result, the probability of the first step in HATI increases dramatically comparing with that of the monochromatic IR laser case; then the ionized electron may absorb more IR photons in LAC process and form the first plateau in the HATI spectrum.

Next, we consider the second plateau shown in Fig.~\ref{fig2}(b). We find that there are two channels (1,1) and (2,0) to make dominate contributions to the second plateau, as shown in Fig.~\ref{fig4}, where these two channels provide comparable contributions to the plateau. Furthermore, it is found that the dip structure only comes from the contribution of channel (1,1), while the angular distribution of channel (2,0) is similar to that of channel (1,0). For the channel (1,1), the electron absorbs another XUV photon in the LAC process; but for channel (2,0), the electron does not absorb any XUV photon in the LAC process. These results indicate that absorbing XUV photon in the LAC process can change the angular distribution of the HATI process, especially can form the dip structure.

In order to explain the interference patterns shown in Fig.~\ref{fig4}, we can consider the HATI as such a process: the ionized electron with certain momentum $\textbf{p}_n$ in the ATI process collides with its parent ion, and then gets its final Volkov state in the IR+XUV two-color laser fields. According to our theory, the interference pattern may be attributed to the coherent summation of the contributions from different subchannels. Hence we further define the subchannel as $(s_2|s_1,q_2)$ within channel $(s_2,q_2)$, where $s_1$ is the number of IR photons absorbed ($s_1>0$) or emitted ($s_1<0$) in the  ATI process.

We now consider the interference pattern of channel (1,1) shown in Fig.~\ref{fig4}(b). Figs.~\ref{fig5}(a)-\ref{fig5}(d) present the channel contributions for channel (1,1) with certain emission angles of $\textbf{p}_n$ at $\theta_n=20^\circ$, $\theta_n=40^\circ$, $\theta_n=65^\circ$ and $\theta_n=160^\circ$, respectively. $\theta_n$ is the angle between the electron momentum $\textbf{p}_n$ and the polarization direction of the laser fields. From Figs.~\ref{fig5}(a)-\ref{fig5}(c), one can find that there exists a narrow dip structure in each graph, and the dip structure moves toward the line defined by $\theta_f=90^\circ$ with the increase of $\theta_n$. Furthermore, comparing Figs.~\ref{fig5}(a) and \ref{fig5}(d), one can see that the two patterns are symmetrical, which leads to the formation of symmetrical structure shown in Fig.~\ref{fig4}(b). In order to further explain interference pattern in more detail, we present the subchannel contributions for (1$|-$9,1) [Figs.~\ref{fig5}(a1)-\ref{fig5}(d1)], (1$|$2,1) [Figs.~\ref{fig5}(a2)-\ref{fig5}(d2)] and  (1$|$12,1) [Figs.~\ref{fig5}(a3)-\ref{fig5}(d3)] with certain emission angles of  $\theta_n=20^\circ$ [Figs.~\ref{fig5}(a1)-\ref{fig5}(a3)], $\theta_n=40^\circ$ [Figs.~\ref{fig5}(b1)-\ref{fig5}(b3)], $\theta_n=65^\circ$ [Figs.~\ref{fig5}(c1)-\ref{fig5}(c3)] and $\theta_n=160^\circ$ [Figs.~\ref{fig5}(d1)-\ref{fig5}(d3)]. It is shown that there is a waist in each of the interference patterns for every subchannel. From Fig.~\ref{fig5}(a1) to \ref{fig5}(a3), one can see that the waist in each graph gradually moves toward the line defined by $\theta_f=0$ with the increase of the number of the IR photons $s_1$ absorbing by the electron in the ATI process. Comparing Fig.~\ref{fig5}(a) with Figs.~\ref{fig5}(a1)-\ref{fig5}(a3), it is found that the interference pattern shown in Fig.~\ref{fig5}(a) comes from the summations of the contributions from different subchannels. Moreover, one can find that the minimum value appears at the waist for subchannel (1$|s_1$,1). Hence these minimum  values of the subchannels lead to the formation of the dip structure, rather than the narrowest distribution around $\theta_f=90^\circ$ as shown in Fig.~\ref{fig5}(a). To further demonstrate the above analysis, we have calculated the angular distributions of different partial subchannels, and find that only the coherent summation of contributions of all subchannels can lead to the narrowest distribution around $\theta_f=90^\circ$ presented in Fig.~\ref{fig5}(a). This tells us that the formation of the narrowest distribution comes from the quantum interference of the contributions of all subchannels, while the dip structure is from the the direct summation over the minimum value of the waists for different subchannels.

Next we consider the interference pattern of channel (2,0) shown in Fig.~\ref{fig4}(c). Similarly, Figs.~\ref{fig6}(a)-\ref{fig6}(d) present the channel contributions for channel (2,0) with $\theta_n=20^\circ$, $\theta_n=40^\circ$, $\theta_n=60^\circ$ and $\theta_n=160^\circ$, respectively. One can see that there is no dip structure whatever direction that the ionized electron emits along with in the ATI process. In order to understand the above phenomena, we also present the subchannel contributions of (2$|-$19,0) [Figs.~\ref{fig6}(a1)-\ref{fig6}(d1])], (2$|-$9,0) [\ref{fig6}(a2)-\ref{fig6}(d2)] and (2$|$2,0) [\ref{fig6}(a3)-\ref{fig6}(d3)], under the corresponding emission directions. One can see that the final kinetic energy of ionized electron increases and ionization probability decreases with absorbing the number of IR photons in the ATI process. In addition, from Figs.~\ref{fig6}(a1)-\ref{fig6}(a3), it is found that the waist of the subchannel does not move and is nearly located at the line defined by $\theta_f=\theta_n$ with the increase of $s_1$. Furthermore, one can find that the  maximum value appears at the waist for subchannel (2$|s_1$,0). These results are obviously different from the interference of the waists for subchannel (1$|s_1$,1). It is the reason why there is no dip structure for channel (2,0), but a dip structure appears for channel (1,1). This phenomenon also tells us that the XUV laser field can change the angular distribution of HATI process and plays a crucial roles for formation of the dip structure in the LAC process.

\subsection {The saddle-point approximation }

In order to understand more deeply the interference patterns of the subchannel angular distribution on the HATI spectrum shown in Fig.~\ref{fig5} and Fig.~\ref{fig6}, we now focus on the analysis of the LAC process by the saddle-point approximation. Under our present calculation condition, the Bessel function in  Eq.~(\ref{7}) can be reduced as
\begin{equation}\label{8}
  \aleph_{q_1 q_2} (\zeta_f-\zeta_n)=J_{q_1}(\zeta_{n_1}-\zeta_{f_1}) J_{q_2}(\zeta_{n_2}-\zeta_{f_2}).
\end{equation}
For absorbing a certain number of XUV photons $q_2$, the Bessel function $J_{q_1}(\zeta_{n_1}-\zeta_{f_1})$ can be expressed in an integral form
\begin{equation}\label{9}
 J_{q_1}(\zeta_{n_1}-\zeta_{f_1})=\frac{\omega_1}{2\pi} \int_{-T_1/2} ^{T_1/2} dt \exp\{i[(\zeta_{n_1}-\zeta_{f_1}) \sin(\omega_1 t)-q_1\omega_1 t]\}, \\
\end{equation}	
where $T_1=2\pi/\omega_1$. On the other hand, the IR laser field can be treated as a classical field, $\textbf{A}_{cl}(t)=\hat{\varepsilon}_1 E_1/\omega_1 \cos(\omega_1 t) $, where $E_1$ is the amplitude of the laser's electric field and $\hat{\varepsilon}_1 $ is the polarization direction. Therefore, the classical action of an electron is
\begin{equation}\label{10}
\begin{array}{l}
S_{cl}(t,\textbf{p}) =  \frac{1}{2}\int_0 ^t dt^{'} [\textbf{p}+e\textbf{A}_{cl}(t^{'})]^2 \\
 ~~~~~~~~~~~=(\frac{1}{2}\textbf{p}^2+U_{p_1})t+2\sqrt {\frac{{{u_{p}}}_1}{{{\omega_1}}}} \sin(\omega_1 t) {{\mathbf{\text{\textbf{p}}}}}\cdot {{\widehat{{\mathbf{\epsilon}}}}_1}+\frac{1}{2}u_{p_1} \sin(2\omega_1 t),\\
 \end{array}
\end{equation}	
where $U_{p_1}=E^2_1/(4\omega_1^2)$ is the ponderomotive energy in the IR laser field. By using Eq.~({\ref{10}}) and the energy conservation $E_i=E_n$ in the ATI process, Eq.~({\ref{9}}) can be cast into
\begin{equation}\label{11}
 J_{q_1}(\zeta_{n_1}-\zeta_{f_1})=\frac{\omega_1}{2\pi} \int_{-T_1/2} ^{T_1/2} dt e^{i f(t)},
\end{equation}	
where $f(t)=S_{cl}(t,\textbf{p}_n)-S_{cl}(t,\textbf{p}_f)+q_2 \omega_2 t$. By using the saddle-point approximation, Eq.~(\ref{11}) can be rewritten as
\begin{equation}\label{12}
 J_{q_1}(\zeta_{n_1}-\zeta_{f_1})=\frac{2 \omega_1}{\sqrt{\pi f''(t_0)}} \cos[F(t_0) -\pi/4],
\end{equation}
where $F(t_0)=(\zeta_{n_1}-\zeta_{f_1}) \sin(\omega_1 t_0)- q_1 \omega_1 t_0$, and the saddle-point $t_0$ satisfies $\cos(\omega_1 t_0)=\frac{q_1}{\zeta_{n_1}-\zeta_{f_1}}=\frac{q_1\omega_1}{2\sqrt{U_{p_1}}({p}_{f\parallel}-{p}_{n\parallel})}$, where ${p}_{f\parallel}={p}_f\cos\theta_f$ and ${p}_{n\parallel}={p}_n\cos \theta_n$. Hence we obtain the energy conservation relationship in the LAC process
\begin{equation}\label{13}
  \frac{1}{2}[\textbf{p}_f+e\textbf{A}(t_0)]^2=\frac{1}{2}[\textbf{p}_n+e\textbf{A}(t_0)]^2+q_2\omega_2,
\end{equation}
where $q_2$ is the number of XUV photons absorbed by the electron. According to Eq.~(\ref{13}), if $q_2=0$, the LAC process is an elastic collision; otherwise, if $q_2\neq0$, the LAC is an inelastic collision process, which indicates that the XUV laser field can change the collision process. Based on Eq.~(\ref{13}), the final energy of ionized electron $E_f$ can be written as
\begin{equation}\label{14}
\begin{array}{l}
E_f=[\sqrt{2U_{p_1}\cos^2(\omega_1 t_0) \cos^2\theta_f+E_n+2\sqrt{2E_nU_{p_1}} \cos(\omega_1 t_0) \cos\theta_n+q_2\omega_2} \\
~~~~~~~-\sqrt{2U_{p_1}} \cos(\omega_1 t_0) \cos\theta_f]^2. \\
 \end{array}
\end{equation}
Here, $E_n=\textbf{p}^2_n/2$ is the energy of the ionized electron in the ATI process. Therefore, the final energy $E_f$ of ionized electron is determined by the above energy orbital equation~(\ref{14}) in the LAC process.

Taking subchannels (1$|-$9,1) and (2$|-$9,0) with $\theta_n=20^\circ$ as examples, Fig.~\ref{fig7} presents the subchannel contributions and corresponding energy orbits for (1$|-$9,1) [Fig.~\ref{fig7}(a)] and (2$|-$9,0) [Fig.~\ref{fig7}(b)]. The insets in Fig.~\ref{fig7} show the details of the waists. One can find that the energy orbits in Fig.~\ref{fig7} agree well with the numerical results, which indicates that the patterns come from the interference between these different orbits. We have noted that the energy of the electron with minimum value at ${\theta_f=0}$ increases with $\theta_f$, while the energy with maximum value at ${\theta_f=0}$ decreases with the angle. Therefore all of these energy orbits intersect at the waist, and the positions of these orbits exchange as $\theta_f$ increases through the waist  from 0 to 180$^\circ$. It tells us that the energy that the ionized electron obtains from the IR laser field in the LAC process varies with the emission angle $\theta_f$. Especially, the ionized electron obtains the maximum energy when it is emitted along the opposition direction of laser polarization. These results indicate that the IR laser field plays an important role in the LAC process. In other worlds, the IR laser field can accelerate or decelerate the electron in the LAC process, which leads to the formation of plateau in HATI process. Furthermore, one can see that the range of energy distribution becomes wider as the number of the IR photons that electron absorbs or emits in the LAC process increases. However, it is noteworthy that there exists corresponding relationship between $q_1=0$ and ${p}_{f\parallel}={p}_{n\parallel}$ at the waist, which indicates that, the momentum of the electron parallel to the laser polarization keeps constant before and after the collision, and also the vector potential of IR laser field is equal to zero in the LAC process. For the waist of the subchannel (1$|-$9,1), the electron absorbs one XUV photon and zero IR photon in the LAC process, but the momenta parallel to the laser polarization before and after collision still remain constant, hence we have $ \frac{1}{2}(p_{f\perp})^2=\frac{1}{2}(p_{n\perp})^2+q_2\omega_2,$ where $p_{f\perp}$ ($p_{n\perp}$) is the momentum of the electron perpendicular to the laser polarization direction after (before) the collision. As we know that the absorbtion of the XUV photon is a multiphoton absorbtion process, hence the electron's momentum may change in all directions by absorbing one XUV photon during the collision. However in the above case, the electron's momentum only changes along the direction perpendicular to the laser polarization, resulting in that the probability of the absorbtion process during the collision gets a minimum value. Hence the probability at the waist becomes minimum value on the spectrum for subchannel (1$|s_1$,1), as shown in the inset in Fig.~\ref{fig7}~(a), leading to the dip structure for channel~(1,1). On the contrary, for the case of subchannel (2$|-$9,0), since electron does not absorb any IR or XUV photon at the waist in the LAC process, its momentum keeps constant before and after the collision, i.e., $\textbf{p}_f=\textbf{p}_n,$. Hence the probability at the waist keeps a certain value for subchannel (2$|s_1$,0), as shown in the inset in Fig.~\ref{fig7}~(b), leading to the fact that there is no dip structure for channel~(2,0).

Furthermore, by using  channel analysis, one can find that the interference pattern shown in Fig.~\ref{fig1} is from the coherent summation of many subchannels. By employing the saddle-point approximation, we may find that the interference pattern is attributed to the term $\cos[F(t_0) -\pi/4]$ in  Eq.~(\ref{12}), which is due to the interference of the orbits at the collision moments $t_0$ and $2\pi/\omega_1-t_0$. When $\cos[F(t_0) -\pi/4]=0$, the destructive interference appears in the angular distribution, as shown by the closed squares in Figs.~\ref{fig9}(a) and \ref{fig9}(b) for subchannels (1$|-$9,1) and (2$|-$9,0) with $\theta_n=20^\circ$ respectively. One can see that the distributions of closed squares agree well with the destructive interference fringes in the spectrum, which indicates that the interference pattern is attributed to the interference of different orbits at collision moments $t_0$ and $2\pi/\omega_1-t_0$ in the LAC process.

\section{CONCLUSIONS}

  Based on the frequency-domain theory, we have investigated the HATI process of an atom in IR+XVU two-color laser fields. It has been shown that the angular distribution presents a multiplateau structure, where the probability of the first plateau is much larger than that of the second plateau and there is a dip structure in the second plateau. In this theory, the HATI can be treated as a two-step process: an ATI followed by an LAC. With the help of channel analysis, it has been found that the angular distribution is attributed to the coherent summation over contributions of different channels, and the dip structure is attributed to the XUV photon absorbtion by the ionized electron in the LAC process. Furthermore, it has been demonstrated that the formation of the narrowest distribution results from the quantum interference of the contributions of all subchannels, while the dip structure is caused by the direct summation over the minimum values at the waists of all subchannels. Moreover, by using the saddle-point approximation, we have obtained an energy equation of classical orbits and have shown that the predictions of the destructive interference positions by the classical energy orbits agrees well with the quantum interference pattern. Additionally, it has been demonstrated that the interference pattern of HATI process comes from the interference of different orbits at collision moments $t_0$ and $2\pi/\omega_1-t_0$ in the LAC process.

\section*{ACKNOWLEDGMENTS}
This research was supported by the National Natural Science Foundation of China under Grant Nos. 11334009, 11425414, 11474348 and 11774411.

\newpage{
\begin{figure}
\includegraphics[width=0.6\textwidth]{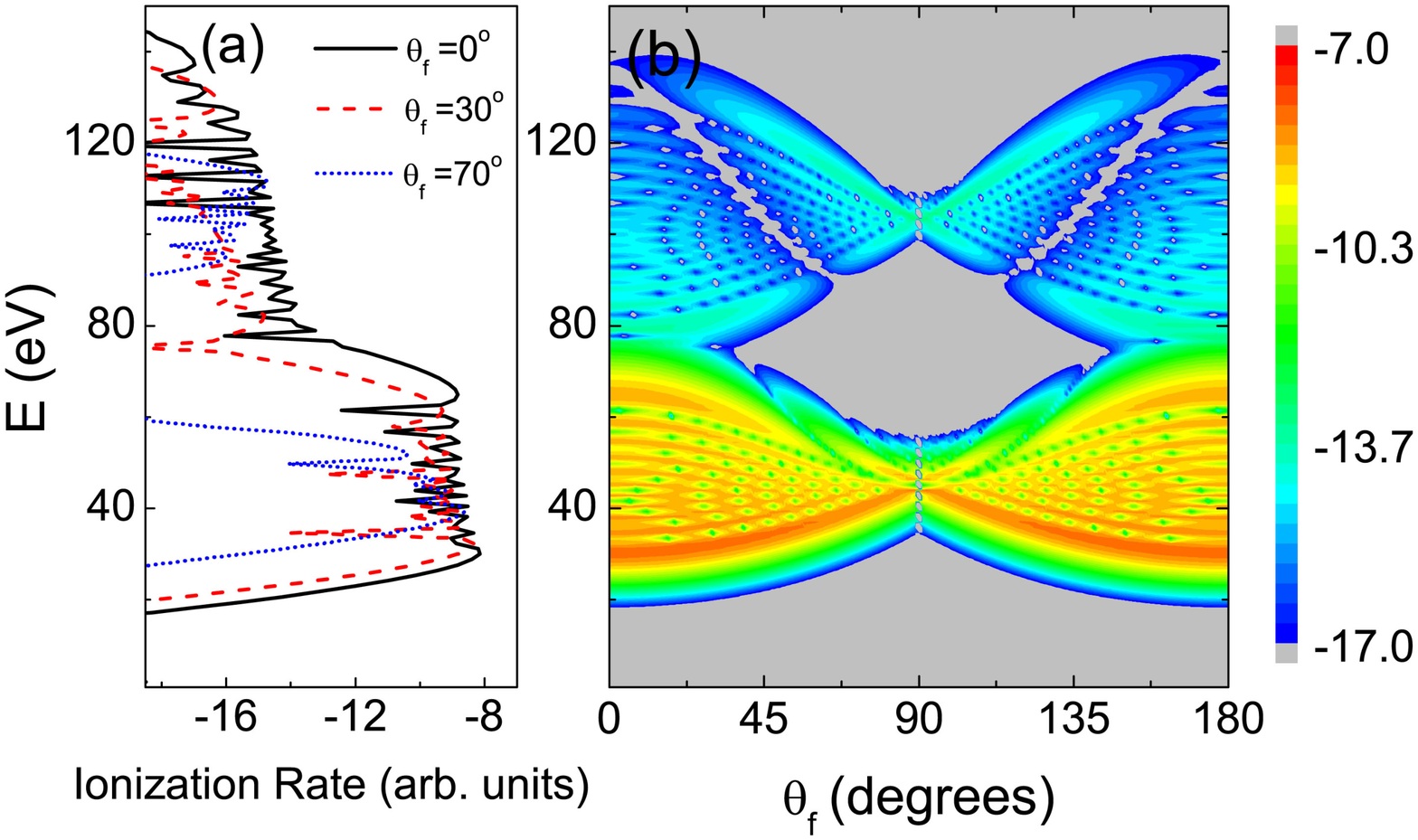}
\caption{(Color online). The angular distribution of an electron in the HATI process, where $\theta_f$ is the angle between the laser polarization and the emission direction of the ionized electron (on logarithmic scale).}
\label{fig1}
\end{figure}}

\newpage{
\begin{figure}
\includegraphics[width=0.6\textwidth]{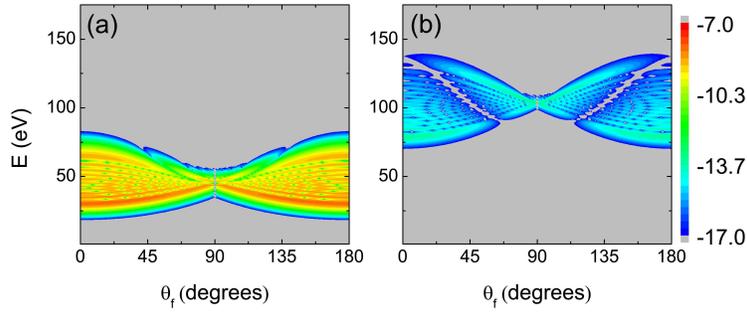}
\caption{(Color online). The angular distribution of the atom absorbing (a) one and (b) two XUV photons in the HATI process (on logarithmic scale).}
\label{fig2}
\end{figure}}

\newpage{
\begin{figure}
\includegraphics[width=0.6\textwidth]{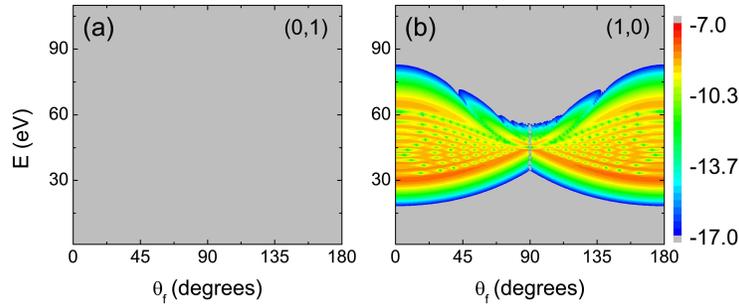}
\caption{(Color online). Channel contributions of the angular distribution for channels (a) (0,1) and (b) (1,0) (on logarithmic scale).}
\label{fig3}
\end{figure}}

\newpage{
\begin{figure}
\includegraphics[width=0.6\textwidth]{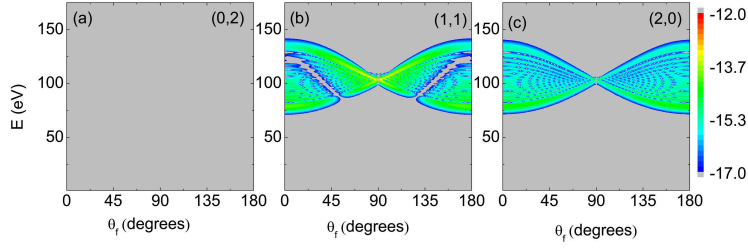}
\caption{(Color online). Channel contributions of the angular distribution for (a) (0,2), (b) (1,1) and (c) (2,0) (on logarithmic scale).}
\label{fig4}
\end{figure}}

\newpage{
\begin{figure}
\includegraphics[width=0.6\textwidth]{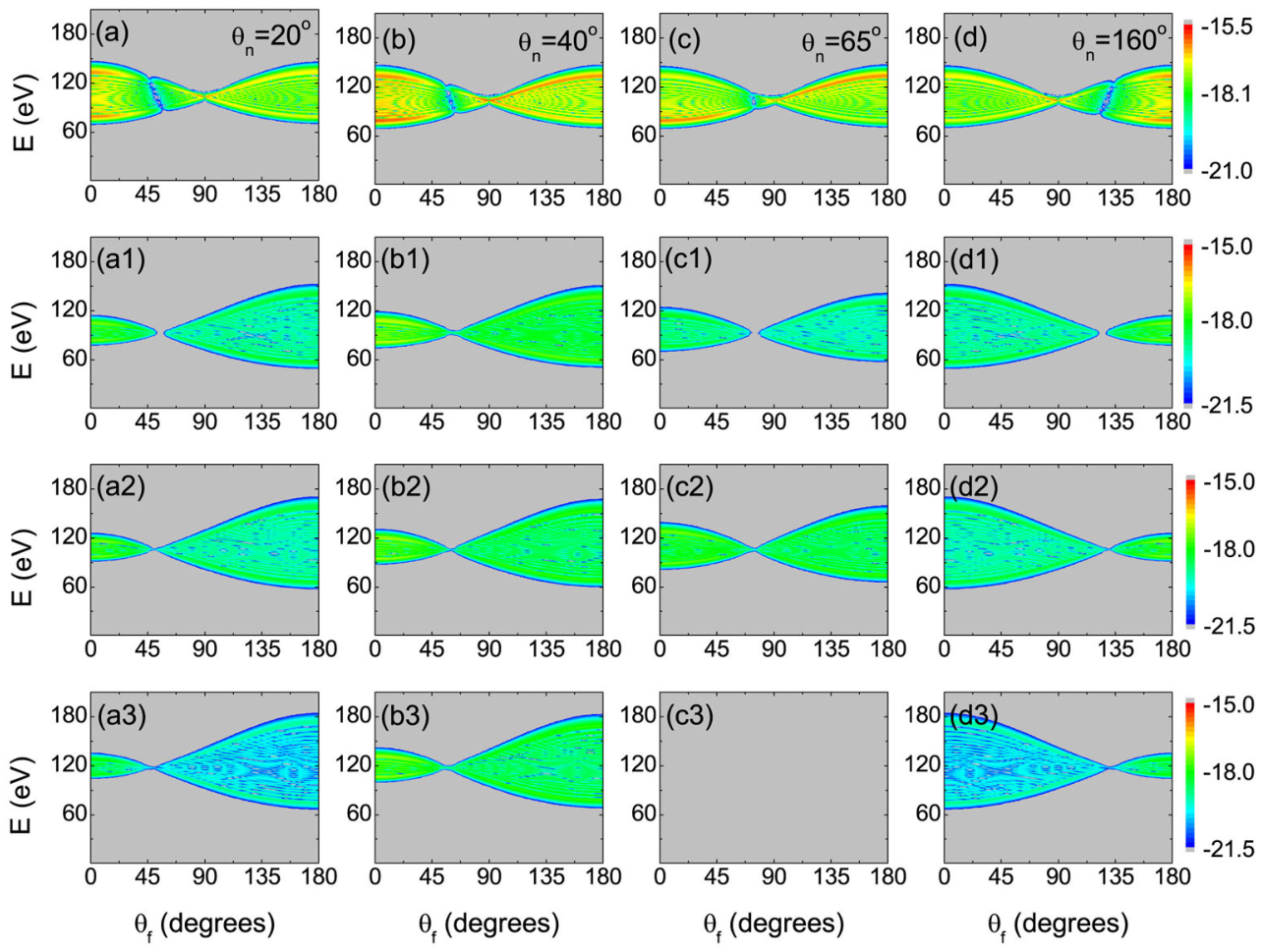}
\caption{(Color online). Channel contributions of the angular distribution for the momentum of the ionized electron $\textbf{p}_n$ in the ATI process along [(a) and (a1)-(a3)] $\theta_n=20^\circ$, [(b) and (b1)-(b3)] $\theta_n=40^\circ$, [(c) and (c1)-(c3)] $\theta_n=65^\circ$ and [(d) and (d1)-(d3)] $\theta_n=160^\circ$. (a)-(d) present channel contributions for channel (1,1). (a1)-(a3), (b1)-(b3), (c1)-(c3) and (d1)-(d3) present subchannel contributions for [(a1)-(d1)] (1$|-$9,1), [(a2)-(d2)] (1$|$2,1) and [(a3)-(d3)] (1$|$12,1) (on logarithmic scale).}
\label{fig5}
\end{figure}}

\newpage{
\begin{figure}
\includegraphics[width=0.6\textwidth]{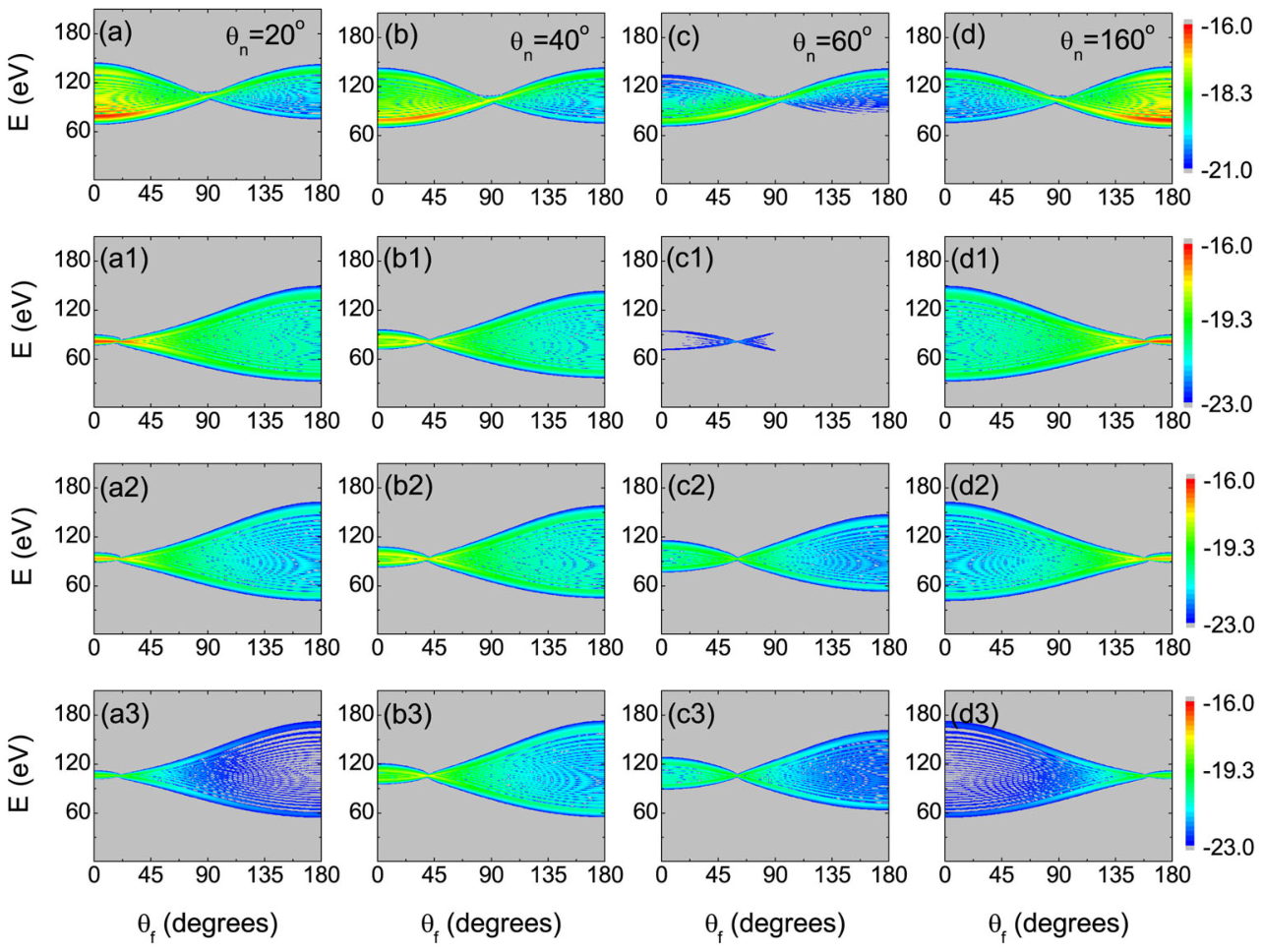}
\caption{(Color online). Channel contributions of the angular distribution for the momentum of the ionized electron $\textbf{p}_n$ in ATI process along [(a) and (a1)-(a3)] $\theta_n=20^\circ$, [(b) and (b1)-(b3)] $\theta_n=40^\circ$, [(c) and (c1)-(c3)] $\theta_n=60^\circ$ and [(d) and (d1)-(d3)] $\theta_n=160^\circ$. (a)-(d) present channel contributions for channel (2,0). (a1)-(a3), (b1)-(b3), (c1)-(c3) and (d1)-(d3) present subchannel contributions for [(a1)-(d1)] (2$|-$19,0), [(a2)-(d2)] (2$|-$9,0) and [(a3)-(d3)] (2$|$2,0) (on logarithmic scale).}
\label{fig6}
\end{figure}}

\newpage{
\begin{figure}
\includegraphics[width=0.6\textwidth]{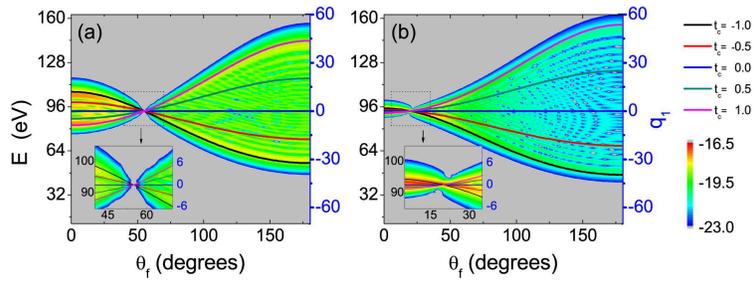}
\caption{(Color online). The final kinetic energy of the ionized electron and the absorbing IR photons $q_1$ in the LAC process versus the emission angle $\theta_f$ for subchannels (a) (1$|-$9,1) and (b) (2$|-$9,0), and the corresponding energy orbits at different collision moments $t_c=\cos(\omega_1 t_0)$ determined by Eq.~(\ref{14}) for $\textbf{p}_n$ along $\theta_n=20^\circ$. The inset presents the details at waist (on logarithmic scale).}
\label{fig7}
\end{figure}}


\newpage{
\begin{figure}
\includegraphics[width=0.6\textwidth]{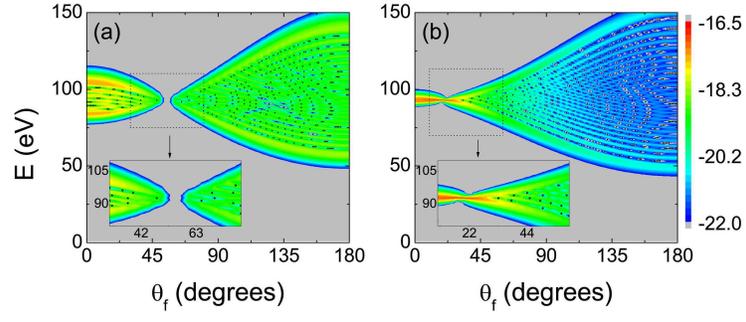}
\caption{(Color online). The channel contributions for subchannels (1$|-$9,1) and (2$|-$9,0), and the closed squares are the locations of destructive interference determined by $\cos[F(t_0)-\pi/4]$ in Eq.~(\ref{12}) for $\textbf{p}_n$ along $\theta_n=20^\circ$. The inset presents the details at waist (on logarithmic scale).}
\label{fig9}
\end{figure}}

\end{document}